\documentclass[preprint,12pt]{elsarticle}

\usepackage{amssymb}
\usepackage{amsmath}

\usepackage{subcaption}
\usepackage{adjustbox}
\usepackage{multirow}
\usepackage[textwidth=3.2cm]{todonotes}

\usepackage[colorlinks=true, linkcolor=blue, citecolor=red, urlcolor=blue]{hyperref}

\usepackage{textgreek}
\usepackage{glossaries}
\glsdisablehyper               
\newacronym{ai}{AI}{Artificial Intelligence}
\newacronym{alr}{ALR}{Adaptive Link Rate}
\newacronym{arm}{ARM}{Advanced RISC Machine}
\newacronym{asic}{ASIC}{Application-Specific Integrated Circuit}
\newacronym{assl}{ASSL}{Adaptive Supply Serial Link}
\newacronym{bts}{BTS}{Back to Sender}
\newacronym{bxi}{BXI}{Bull eXascale Interconnect}
\newacronym{bxiv3}{BXIv3}{Bull eXascale Interconnect version 3}
\newacronym{ca}{CA}{Channel Adapter}
\newacronym{capex}{CAPEX}{Capital Expenditures}
\newacronym{cdf}{CDF}{Cumulative Distribution Function}
\newacronym{cea}{CEA}{Commissariat à l'Énergie Atomique et aux Énergies Alternatives}
\newacronym{cern}{CERN}{European Organization for Nuclear Research}
\newacronym{ci}{CI}{Congestion Isolation}
\newacronym{cicd}{CI/CD}{Continious Delivery/Continious integration}
\newacronym{cioq}{CIOQ}{Combined Input-Output Queued}
\newacronym{cpu}{CPU}{Central Processing Unit}
\newacronym{dbbm}{DBBM}{Destination-Based Buffer Management}
\newacronym{dc}{DC}{Data Center}
\newacronym{dcqcn}{DCQCN}{Data Center Quantized Congestion Notification}
\newacronym{dctcp}{DCTCP}{Data Center TCP}
\newacronym{diffserv}{DiffServ}{Differentiated Services}
\newacronym{dor}{DOR}{Dimensional-order routing}
\newacronym{dvfs}{DVFS}{Dynamic Voltage and Frequency Scaling}
\newacronym{dvl}{DVL}{Dynamic Virtual Lanes}
\newacronym{ecn}{ECN}{Explicit Congestion Notification}
\newacronym{eee}{EEE}{Energy Efficient Ethernet}
\newacronym{eniac}{ENIAC}{Electronic Numerical Integrator and Computer}
\newacronym{fct}{FCT}{Flow Completion Time}
\newacronym{fft}{FFT}{Fast Fourier Transform}
\newacronym{flit}{flit}{flow control unit}
\newacronym{flow2sl}{Flow2SL}{Flow to Service Level}
\newacronym{fpga}{FPGA}{Field-Programmable Gate Array}
\newacronym{gromacs}{GROMACS}{GROningen MAchine for Chemical Simulations}
\newacronym{hol}{HoL}{Head-of-Line}
\newacronym{hoti}{HOTI}{Hot Interconnects}
\newacronym{hpc}{HPC}{High-performance Computing}
\newacronym{http}{HTTP}{HyperText Transfer Protocol}
\newacronym{ia}{IC}{Input Adapter}
\newacronym{ib}{IB}{InfiniBand}
\newacronym{ibta}{IBTA}{InfiniBand Trade Association}
\newacronym{ieee}{IEEE}{Institute of Electrical and Electronics Engineers}
\newacronym{iodet}{IODET}{In-Order DETerministic routing}
\newacronym{iq}{IQ}{Input-Queued}
\newacronym{it}{IT}{information technology}
\newacronym{jos}{JoS}{Journal of Supercomputing}
\newacronym{jsa}{JSA}{Journal of Systems Architecture}
\newacronym{lammps}{LAMMPS}{Large-scale Atomic/Molecular Massively Parallel Simulator}
\newacronym{lan}{LAN}{Local Area Network}
\newacronym{llm}{LLM}{Large Language Model}
\newacronym{lpi}{LPI}{Low Power Idle}
\newacronym{mac}{MAC}{Media Access Control}
\newacronym{man}{MAN}{Metropolitan Area Network}
\newacronym{mlwf}{MLWF}{Machine Learning Weather Forecast}
\newacronym{mpi}{MPI}{Message Passing Interface}
\newacronym{mtu}{MTU}{Maximum Transmission Unit}
\newacronym{nfs}{NFS}{Network File System}
\newacronym{nic}{NIC}{Network Interface Card}
\newacronym{opa}{OPA}{OmniPath}
\newacronym{opex}{OPEX}{Operational Expenditures}
\newacronym{patmos}{PATMOS}{Particle Transport in Modular Systems}
\newacronym{pcie}{PCIe}{PCI Express}
\newacronym{pdt}{PDT}{Power-down Threshold}
\newacronym{pfc}{PFC}{Priority-based Flow Control}
\newacronym{pgas}{PGAS}{Partitioned Global Address Space}
\newacronym{pgft}{PGFT}{Parallel Port Generalized Fat Trees}
\newacronym{phit}{phit}{physical transfer unit}
\newacronym{qos}{QoS}{Quality of Service}
\newacronym{rdma}{RDMA}{Remote Direct Memory Access}
\newacronym{redsea}{RED-SEA}{Network Solutions for Exascale Architectures}
\newacronym{rlft}{RLFT}{Recursive Level Fat-Tree}
\newacronym{roce}{RoCE}{RDMA over Converged Ethernet}
\newacronym{rr}{RR}{Round-Robin}
\newacronym{rtt}{RTT}{Round-Trip Time}
\newacronym{saas}{SaaS}{Software-as-a-Service}
\newacronym{sc}{SC}{Supercomputer}
\newacronym{sfc}{SFC}{Source Flow Control}
\newacronym{sla}{SLA}{Service Level Agreement}
\newacronym{sl}{SL}{Service Level}
\newacronym{slurm}{SLURM}{Simple Linux Utility for Resource Management}
\newacronym{sqs}{SQS}{Static Queueing Scheme}
\newacronym{tcp}{TCP}{Transmission Control Protocol}
\newacronym{vct}{VCT}{Virtual Cut-Through}
\newacronym{vc}{VC}{Virtual Channel}
\newacronym{vl}{VL}{Virtual Lane}
\newacronym{vm}{VM}{Virtual Machine}
\newacronym{voq}{VOQ}{Virtual Output Queue}
\newacronym{voqnet}{VOQnet}{Virtual Output Queues at network level}
\newacronym{voqsw}{VOQsw}{Virtual Output Queues at the switch level}
\newacronym{vrm}{VRM}{Voltage Regulation Module}
\newacronym{wan}{WAN}{Wide Area Network}

\journal{Journal of Systems Architecture}

\begin{document}

\begin{frontmatter}



\title{Combined power management and congestion control in High-Speed Ethernet-based
Networks for Supercomputers and Data Centers}

\author[uclm]{Miguel Sánchez de la Rosa}  
\author[uva]{Francisco J. Andújar}  
\author[uclm]{Jesus Escudero-Sahuquillo}  
\author[uclm]{José L. Sánchez}  
\author[uclm]{Francisco J. Alfaro-Cortés}  
\affiliation[uclm]{organization={Department of Computing Systems, Universidad de Castilla-La Mancha},
            addressline={Avda. de España},
            city={Albacete},
            postcode={02071},
            state={Castilla-La Mancha},
            country={Spain}}

\affiliation[uva]{organization={Departamento de Informática, Universidad de Valladolid},
            addressline={Plaza de Santa Cruz, 8},
            city={Valladolid},
            postcode={47002},
            state={Castilla y León},
            country={Spain}}



\begin{abstract}
The demand for computer in our daily lives has led to the proliferation of \glspl{dc} that power indispensable many services. On the other hand, computing has become essential for some research for various scientific fields, that require \glspl{sc} with vast computing capabilities to produce results in reasonable time. The scale and complexity of these systems, compared to our day-to-day devices, are like comparing a cell to a living organism. To make them work properly, we need state-of-the-art technology and engineering, not just raw resources. Interconnecting the different computer nodes that make up a whole is a delicate task, as it can become the bottleneck for the whole infrastructure. In this work, we explore two aspects of the network: how to prevent degradation under heavy use with congestion control, and how to save energy when idle with power management; and how the two may interact.

\end{abstract}




\end{frontmatter}
\pagebreak
\section{Introduction}
Energy saving in interconnection networks has often been overlooked, specially in \gls{hpc} environment, because it may lead to performance degradation. However, over the years, as \glspl{cpu} in computing nodes have become more energy-proportional, it has taken more of the power budget for the whole system~\cite{energyproportional}. Links power draw is the same regardless of network activity. In other words, if there are energy-saving opportunities for \glspl{cpu}~\cite{intelsandybridge} and accelerators when idling, thus it would be beneficial to find analogous ones for the network. Furthermore, applications running in \gls{hpc} environments usually alternate between computing and communication intervals, which may not overlap. Lowering the network to a \gls{lpi} state has an execution and latency overhead during which transmission is not possible, and so does getting it back to operation. However, if computing intervals and spaced out and long enough, the concession to allow some performance degradation in ecxchange for power saving can be beneficial. \gls{ib} and Ethernet, the two most common interconnection technologies, have power-saving in \gls{ibta}~\cite{dickov_software-managed_2014} and \gls{eee}~\cite{christensen_ieee_2010}, their respective standards. However, they are often disregarded seeking for the maximum performance possible.

When the network is under heavy load, we face other problems, mainly with arbitration on shared resources. Shared resources are a key aspect of switched networks, but when certain traffic (hot flows) monopolize the usage of links or packet buffers, other communications (victin/cold flows) are subject to serious degradation even if arbitration is still fair. Packets stall on output buffers causing contention, and the result is that the network becomes congested and cannot handle all the traffic that is being injected. Then, congestion control mechanisms like stop-and-go or credit-based for lossless network; or retransmissions in lossy ones, take place. This causes congestion to back-propagate to the sources, drastically degrading overall performance. This phenomenom is known as \gls{hol}.

\section{Related work}
For power management, several proposals based on \gls{eee} have been carried out. The key aspect is when to power down ports after there are no packets to transmit. Thus, a \gls{pdt} \cite{saravanan_powerperformance_2013} was proposed to leave links operational for a constant amount of time after transmission. This prevents degradation on spaced-out packets to be transmitted within that time frame. The main drawback of using a \gls{pdt} is that it is either uninformed about link usage (all network links have the same value regarless of their behavior), or requires fine tuning by executing the workloads beforehand; which is obviously energy-consuming. To solve this, Saravanan~\cite{perfbound} proposed PerfBound, which dynamically calculates new \gls{pdt} values for every port with a set degradation limit.
We proposed an enhancement on PerfBound~\cite{perfboundcorrect} to reduce the degradation, also exploring several strategies to manage the data structure that governs the mechanism.

Congestion control aims to reduce or eliminate \gls{hol} in many different ways. The most straightforward, requiring no hardware modifications is using \glspl{sqs}~\cite{yebenes_efficient_2015} to spread the congestion among different \glspl{vc}. This can be complemented by reducing the \glspl{vc} available for the queuing schemes and devoting specific ones for congesting flows. This is known as \glspl{dvl}~\cite{gonzalez-naharro_efficient_2019,dvl-lossy}. Flow control can also help with congestion, as \gls{pfc}~\cite{pfc} or credit-based mechanisms work on a \gls{vc} instead of the whole link. End-to-end flow control mechanisms can help by completely halting injection in the case of \gls{sfc}~\cite{sfc,sfcintel} or throttling (reducing) congestion flows like \gls{dcqcn}~\cite{pfc} and \gls{dctcp}~\cite{dctcp}.

\section{Experiment results and outcomes}
For our experiments, we have used 4 \glspl{sqs}: 1Q (one queue), BBQ~\cite{segura_bbq_2013}, \gls{dbbm}~\cite{nachiondo_buffer_2010} and \gls{flow2sl}~\cite{yebenes_efficient_2015}. 

\begin{figure}[t]
    \centering
    \begin{subfigure}{0.48\textwidth}
        \includegraphics[width=\linewidth]{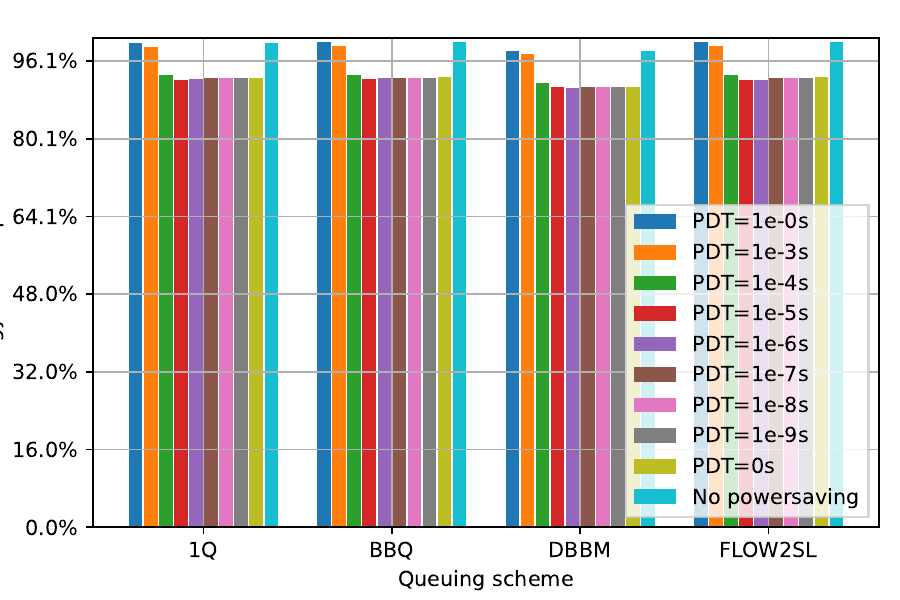}
        \caption{PDT}
        \label{fig:pdt-energy}
    \end{subfigure}
    \hfill
    \begin{subfigure}{0.48\textwidth}
        \includegraphics[width=\linewidth]{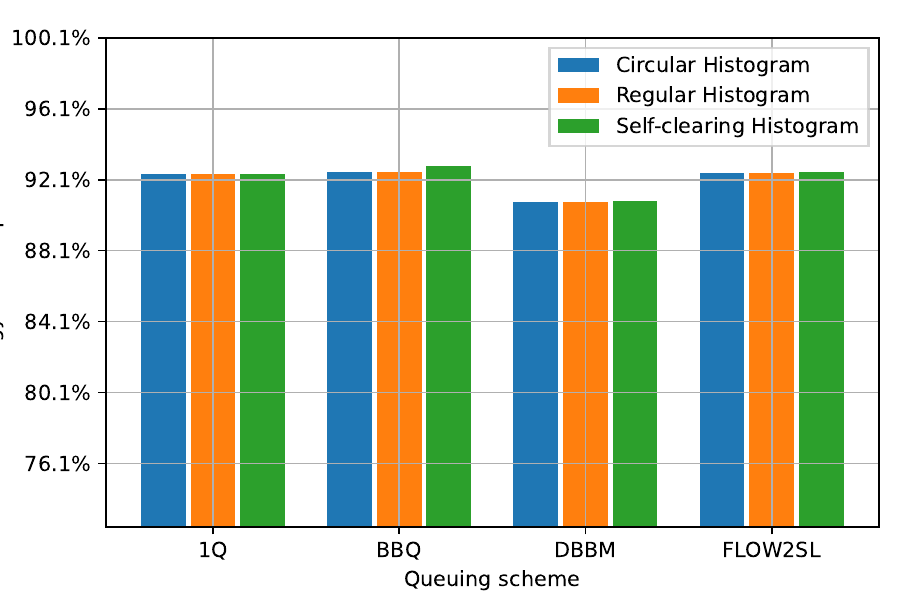}
        \caption{PBOUND}
        \label{fig:pbound-energy}
    \end{subfigure}
    \caption{Energy consumed.}
    \label{fig:energy}
\end{figure}

In \figurename~\ref{fig:energy} we show the energy effects of employing fixed \gls{pdt} values and using PerfBound. We can see subtle differences in energy consume depending on the \gls{sqs} employed (\figurename~\ref{fig:pdt-energy}), but the driving factor in energy is the \gls{pdt} employed. Using PerfBound, as shown in \figurename~\ref{fig:pbound-energy} further reduces energy consumption, and complements the \gls{dbbm} \gls{sqs}.

\begin{figure}[t]
    \centering
    \begin{subfigure}{0.48\textwidth}
        \includegraphics[width=\linewidth]{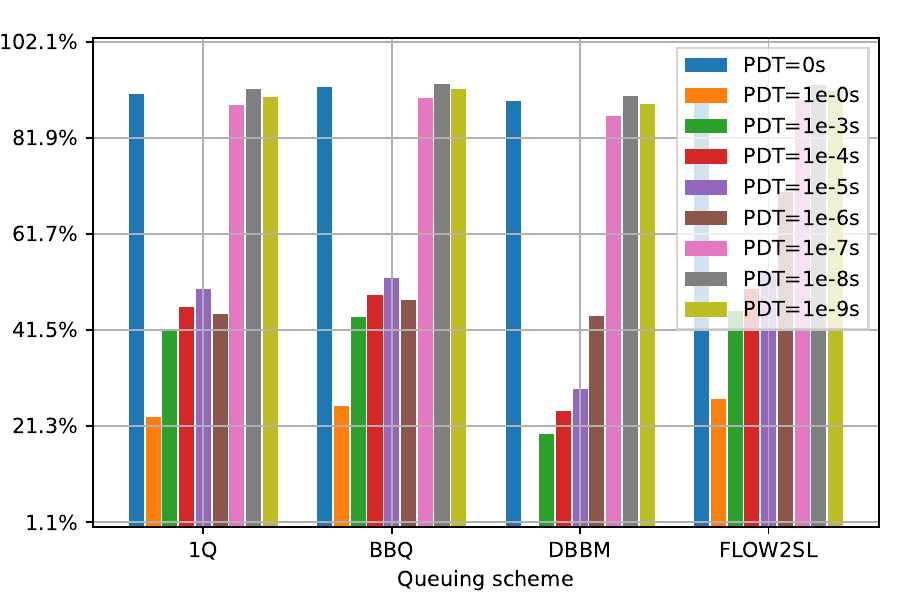}
        \caption{PDT}
        \label{fig:pdt-netlat}
    \end{subfigure}
    \hfill
    \begin{subfigure}{0.48\textwidth}
        \includegraphics[width=\linewidth]{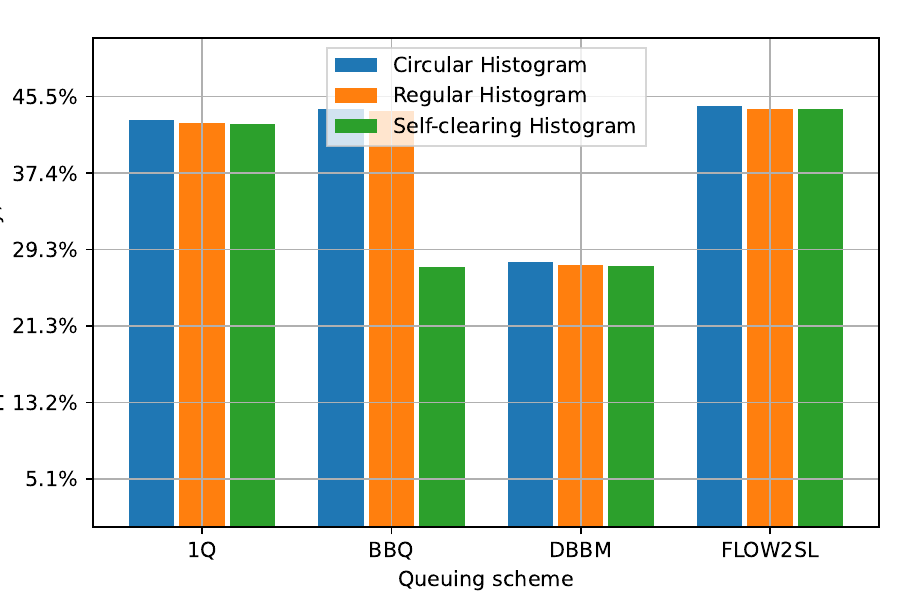}
        \caption{PBOUND}
        \label{fig:pbound-netlat}
    \end{subfigure}
    \caption{Net Latency increase.}
    \label{fig:netlat}
\end{figure}

If we look at the network latency, we can see in \figurename~\ref{fig:netlat} that differences start to show with different PDT values (see \figurename~\ref{fig:pdt-netlat}). As shown in \figurename~\ref{fig:pbound-netlat}, differences between \glspl{sqs} become more apparent when using PerfBound, even if the histogram management strategy produces minimal changes.

\begin{figure}[t]
    \centering
    \begin{subfigure}{0.48\textwidth}
        \includegraphics[width=\linewidth]{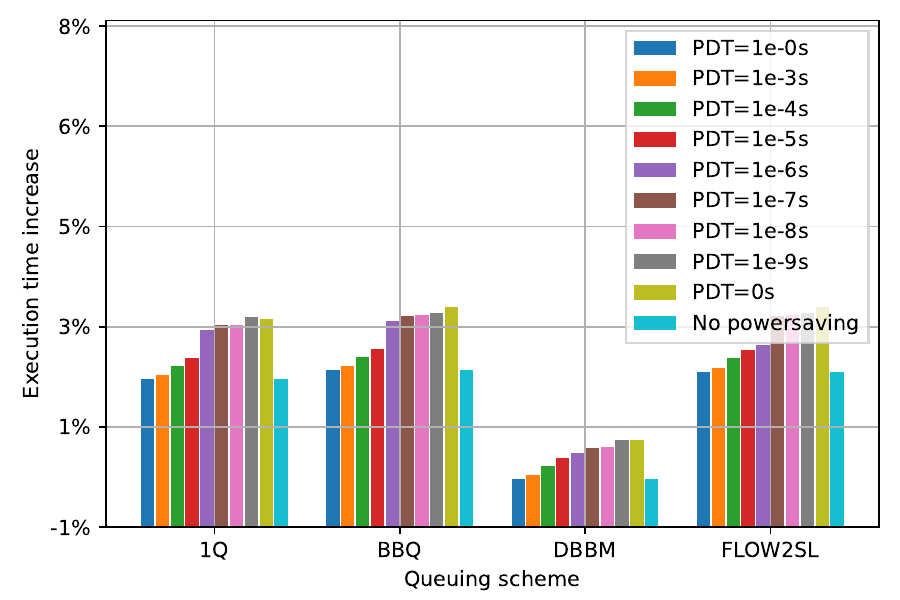}
        \caption{PDT}
        \label{fig:pdt-runtime}
    \end{subfigure}
    \hfill
    \begin{subfigure}{0.48\textwidth}
        \includegraphics[width=\linewidth]{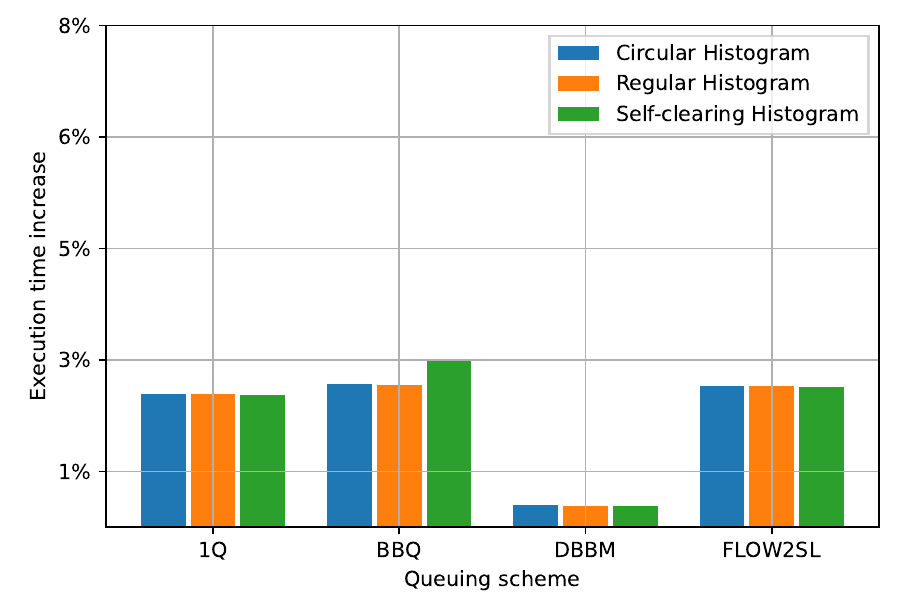}
        \caption{PBOUND}
        \label{fig:pbound-runtime}
    \end{subfigure}
    \caption{Runtime increase.}
    \label{fig:runtime}
\end{figure}

In \figurename~\ref{fig:runtime} we show the execution time increase when using the different \glspl{sqs} and power management techniques. In \figurename~\ref{fig:pdt-runtime} we see that all SQSs have increasing overhead values as the PDT becomes more restrictive, but DBBM is significantly less degrading. In \figurename~\ref{fig:pbound-runtime}, we see that the effect of SQSs when using PerfBound is even more evident.

In this work, we have highlighted the importance of combining SQSs with two well-regarded power management techniques. In particular, we have seen how queuing schemes have net benefitial effects on both power consumption and reduced overhead on both latencies and execution time.

\paragraph{Acknowledgements}
This work has been supported by the Junta de Comunidades de Castilla-La Mancha under projects SBPLY/21/180225/000103, SBPLY/21/180501/000248 and TED2021-130233B-C31, and by the Spanish Ministry of Science and Innovation (MCIN) / Agencia Estatal de Investigación (AEI) under project PID2021-123627OB-C52, co-financed by the European Regional Development Fund (FEDER). We also acknowledge support from the PERTE-Chip grants (UCLM Chair, TSI-069100-2023-0014) from the Spanish Ministry of Digital Transformation and Public Service.




\bibliographystyle{plain}  
\bibliography{ref}  
\end{document}